\documentclass{aa}  
\usepackage[dvipsnames]{xcolor}
\usepackage{graphicx}
\usepackage{multirow}
\usepackage[shortlabels]{enumitem}
\usepackage{ulem}
\usepackage{txfonts}
\usepackage{subfigure}
\usepackage[
  breaklinks = true,
  colorlinks = true,
  urlcolor   = blue,
  citecolor  = blue,
  linkcolor  = blue,
]{hyperref}
\graphicspath{ {./figures/} }

\newcommand{\IRIS}{IRIS}
\newcommand{\SDO}{SDO}
\newcommand{\Halpha}{{H$\alpha$}}
\newcommand{\Hbeta}{{H$\beta$}}
\newcommand{\CaII}{{\ion{Ca}{II}}}
\newcommand{\CaIIK}{{\ion{Ca}{II}~K}}
\newcommand{\CaIIH}{{\ion{Ca}{II}~H}}

\newcommand{\FeI}{{\ion{Fe}{I}}}

\newcommand{\MgII}{{\ion{Mg}{II}}}

\newcommand{\FeXII}{{\ion{Fe}{XII}}}

\begin{document} 
%---------------------------------------------------------------------------------------------------------------------------
   \title{Small-scale magnetic flux emergence preceding a chain of energetic solar atmospheric events}
   \titlerunning{Small-scale magnetic flux emergence}

%---------------------------------------------------------------------------------------------------------------------------
   \author{D.~N\'obrega-Siverio\inst{1,2,3,4}
          \and
          I.~Cabello\inst{1,2,5}
          \and
          S. Bose\inst{6,7,3,4}
          \and 
          L.~H.~M.~Rouppe~van~der~Voort\inst{3,4}
          \and \\
          R. Joshi\inst{3,4}
          \and
          C.~Froment\inst{8}
          \and
          V.~M.~J.~Henriques\inst{3,4}
          }
          
%---------------------------------------------------------------------------------------------------------------------------          
   \institute{Instituto de Astrof\'isica de Canarias, E-38205 La Laguna, Tenerife, Spain\\
    \email{dnobrega@iac.es}
    \and
    Universidad de La Laguna, Dept. Astrof\'isica, E-38206 La  Laguna, Tenerife, Spain
    \and
    Rosseland Centre for Solar Physics, University of Oslo, PO Box 1029 Blindern, 0315 Oslo, Norway
    \and
    Institute of Theoretical Astrophysics, University of Oslo, PO Box 1029 Blindern, 0315 Oslo, Norway
    \and
    Universidad de La Laguna, Dept. Did\'acticas Espec\'ificas, E-38200 
    La Laguna, Tenerife, Spain
    \and
    Lockheed Martin Solar and Astrophysics Laboratory, Palo Alto, CA 94304, USA
    \and
    Bay Area Environmental Research Institute, NASA Research Park, Moffett Field, CA 94035, USA
    \and
    LPC2E, CNRS/CNES/University of Orl\'eans, 3A avenue de la Recherche Scientifique, Orl\'eans, France}

%---------------------------------------------------------------------------------------------------------------------------
   \date{Received December 11, 2023; accepted March 17, 2024}

%%%%%%%%%%%%%%%%%%%%%%%%%%%%%%%%%%%%%%%%%%%%%%%%%%%%%%%%%%%%%%%%%%%%%%%%%%%%%%%%%%%%%%%%%%%%%%%%%%%%%%%%%%%%%%%%%%%%%%%%%%%%
% ABSTRACT
%%%%%%%%%%%%%%%%%%%%%%%%%%%%%%%%%%%%%%%%%%%%%%%%%%%%%%%%%%%%%%%%%%%%%%%%%%%%%%%%%%%%%%%%%%%%%%%%%%%%%%%%%%%%%%%%%%%%%%%%%%%%
\abstract
%---------------------------------------------------------------------------------------------------------------------------
% Context
{Advancements in instrumentation have revealed a multitude of small-scale extreme-ultraviolet (EUV) events in the solar atmosphere
and considerable effort is currently undergoing to unravel them.}
%---------------------------------------------------------------------------------------------------------------------------
% Aims 
{Our aim is to employ high-resolution and high-sensitivity magnetograms to gain a detailed understanding of the magnetic origin of such phenomena.}
%---------------------------------------------------------------------------------------------------------------------------
% Methods
{We have used coordinated observations from the Swedish 1-m Solar Telescope (SST), the Interface Region Imaging 
Spectrograph (\IRIS), and the Solar Dynamics Observatory (\SDO) to analyze an ephemeral 
magnetic flux emergence episode and the following chain of small-scale energetic events.
These unique observations clearly link these phenomena together.
}
%---------------------------------------------------------------------------------------------------------------------------
% Results
{The high-resolution (0\farcs057 pixel$^{-1}$) magnetograms obtained with SST/CRISP allows us to reliably measure the 
magnetic field at the photosphere and detect the emerging bipole that causes the subsequent eruptive atmospheric events.
Notably, this small-scale emergence episode remains indiscernible in the lower 
resolution SDO/HMI magnetograms (0\farcs5 pixel$^{-1}$). 
We report the appearance of a dark bubble in \CaIIK\ 3933 \AA\ related to the emerging bipole, a sign 
of the canonical expanding magnetic dome predicted in flux emergence simulations.
Evidences of reconnection are also found: first through an Ellerman bomb, and later by the launch of a surge next to a UV burst.
The UV burst exhibits a weak EUV counterpart in the coronal SDO/AIA channels. 
By calculating the differential emission measure (DEM), its plasma is shown to reach a temperature beyond 1 MK and have densities between the upper chromosphere and transition region.}
%---------------------------------------------------------------------------------------------------------------------------
% Conclusions
{Our study showcases the importance of high-resolution magnetograms to unveil the mechanisms triggering phenomena such as EBs, UV bursts, and surges.
This could hold implications for small-scale events akin to those recently reported in EUV using Solar Orbiter.
The finding of temperatures beyond 1 MK in the UV burst plasma strongly suggests that we are examining analogous features.
Therefore, we signal caution regarding drawing conclusions from full-disk magnetograms that lack the necessary resolution to reveal their true magnetic origin.}

%---------------------------------------------------------------------------------------------------------------------------
% Keywords
\maketitle

%%%%%%%%%%%%%%%%%%%%%%%%%%%%%%%%%%%%%%%%%%%%%%%%%%%%%%%%%%%%%%%%%%%%%%%%%%%%%%%%%%%%%%%%%%%%%%%%%%%%%%%%%%%%%%%%%%%%%%%%%%%%
% SECTION 1: INTRODUCTION
%%%%%%%%%%%%%%%%%%%%%%%%%%%%%%%%%%%%%%%%%%%%%%%%%%%%%%%%%%%%%%%%%%%%%%%%%%%%%%%%%%%%%%%%%%%%%%%%%%%%%%%%%%%%%%%%%%%%%%%%%%%%
\section{Introduction}
%---------------------------------------------------------------------------------------------------------------------------
Magnetic flux emergence from the solar interior into the photosphere and upper layers is key to triggering  a wide variety 
of eruptive and ejective phenomena in the solar atmosphere. 
From observations, there are several reports about the subsequent phenomena following magnetic flux emergence:
\begin{itemize}
    \item Ellerman bombs \citep[EBs, e.g.,][]{Pariat_etal:2004,Hashimoto:2010,Watanabe_etal:2011,Nelson_etal:2013,Rutten_etal:2013,Vissers_etal:2013,Reid_etal:2016,
    Yang_etal:2016,Rouppe-van-der-Voort_etal:2021,Shen_etal:2022,Rouppe-van-der-Voort_etal:2023};
    %Libbrecht_etal:2017,   %% no related to flux emergence (Close to the limb)
    %JoshiJayant_etal:2020, %% no related to flux emergence (QS)
    %JoshiJayant_etal:2022, %% no related to flux emergence (QS)
    \item surges \citep[e.g.,][]{Liu_Kurokawa:2004,Guglielmino_etal:2010,Vargas-Dominguez_etal:2014,Kim_etal:2015,
    Nobrega-Siverio_etal:2017,Guglielmino_etal:2019,Verma_etal:2020,Diaz-Baso_etal:2021,Nobrega-Siverio_etal:2021};
    \item Ultraviolet (UV) bursts \cite[e.g.,][]{Peter_etal:2014,Vissers_etal:2015,Grubecka_etal:2016,
    Rouppe-van-der-Voort_etal:2017,Young_etal:2018,Tian_etal:2018,Guglielmino_etal:2018,Chen_etal:2019,Ortiz_etal:2020};
    \item and coronal jets \cite[e.g.,][]{Shibata_etal:1992,Canfield_etal:1996,Liu_etal:2011,
    Huang_etal:2012,Schmieder_etal:2013,Ruan_etal:2019,JoshiReetika_etal:2020}. 
\end{itemize}

%---------------------------------------------------------------------------------------------------------------------------
Recently, improvements in observational capabilities, such as the Extreme 
Ultraviolet Imager of the High Resolution Imager \citep[EUI-HRI;][]{Rochus_etal:2020} on board Solar Orbiter 
\citep[SolO;][]{Muller_etal:2020}, have unveiled a forest of smallest-scale events visible in the corona.
This includes the so-called \textit{campfires} \citep[e.g.,][]{Berghmans_etal:2021,Zhukov_etal:2021,Kahil_etal:2022,Dolliou_etal:2023,Huang_etal:2023}, which are small-scale EUV brightenings in the quiet Sun, and small inverted-Y shaped jets with base-width down to $\leq$ 1~Mm \citep{Mandal_etal:2022,Chitta_etal:2023,Panesar_etal:2023}.
In order to broaden our comprehension of their origins, it is crucial to understand the connection to the topology and evolution of the magnetic field. 
This requires coordination of the coronal observations with observations at other wavelengths and spectropolarimetry to cover the full solar atmosphere.  
High-sensitivity in magnetograms is essential for revealing the appearance of emerging small-scale magnetic loops and accurately characterizing the magnetic flux.
For example, \cite{Guglielmino_etal:2018} found that the magnetic flux measured in an emerging region with the 
\textit{Helioseismic and Magnetic Imager} \citep[HMI;][]{Scherrer_etal:2012} onboard the Solar Dynamics Observatory 
\citep[SDO;][]{Pesnell_etal:2012} is three times smaller than the flux measured with the higher resolution 
\textit{Solar Optical Telescope} \citep[SOT;][]{Tsuneta_etal:2008} onboard Hinode \citep{Kosugi_etal:2007}.
Furthermore, high spatial resolution, in observations that sample the lower atmosphere, is 
imperative to be able to detect the rise 
of the small-scale loops through the atmosphere and their interaction with the preexisting ambient field.
For instance, using HINODE/SOT, \cite{Vargas-Dominguez_etal:2012} and \cite{Kontogiannis_etal:2020} identified elliptical 
dark features in \CaIIH\ images with a lifetime of $\approx$12~min located between the footpoints of emerging bipoles, 
a finding also reported by \cite{Centeno_etal:2017} through \CaIIH\  observations with SUNRISE II 
\citep{Solanki_etal:2017}.
Similar features visible in \CaII\ 8542 \AA\ were described by \cite{Ortiz_etal:2014,de-la-Cruz-Rodriguez_etal:2015b,Ortiz_etal:2016}, using the Swedish $1-$m Solar 
Telescope \citep[SST;][]{Scharmer_etal:2003}. 
Further evidence using other spectral lines is necessary to better constrain the magnetic flux emergence 
through the lower atmosphere.

%---------------------------------------------------------------------------------------------------------------------------
In this work, we present a small-scale magnetic flux emergence episode observed on 2017 May 25 using SST, 
IRIS \citep{De-Pontieu_etal:2014}, 
and SDO, underscoring the absolute necessity of sensitive magnetograms for detecting such events.
The episode exhibited a wide variety of signatures in all the atmospheric layers resulting from magnetic flux emergence 
and subsequent reconnection with the preexisting ambient field.
The layout of this paper is as follows. In Sect.~\ref{s:observations}, we describe the coordinated observations and
the alignment between them. 
In Sect.~\ref{s:results}, we show the results, analyzing the magnetic flux emergence episode through different 
resolution magnetograms (Sect.~\ref{s:flux_emergence}). 
We later describe all the phenomena related to it, namely, the dark bubble 
(Sect. \ref{s:dark_bubble}), the Ellerman bomb (Sect. \ref{s:ellerman_bomb}), 
as well as UV burst and surge along with their EUV counterparts 
(Sects.~ \ref{s:uv_burst} and \ref{s:surge}).
Finally, Sect.~\ref{s:discussion} contains the discussion and the main conclusions.

%%%%%%%%%%%%%%%%%%%%%%%%%%%%%%%%%%%%%%%%%%%%%%%%%%%%%%%%%%%%%%%%%%%%%%%%%%%%%%%%%%%%%%%%%%%%%%%%%%%%%%%%%%%%%%%%%%%%%%%%%%%%
% SECTION 2: OBSERVATIONS
%%%%%%%%%%%%%%%%%%%%%%%%%%%%%%%%%%%%%%%%%%%%%%%%%%%%%%%%%%%%%%%%%%%%%%%%%%%%%%%%%%%%%%%%%%%%%%%%%%%%%%%%%%%%%%%%%%%%%%%%%%%%
\section{Observations}\label{s:observations}
%---------------------------------------------------------------------------------------------------------------------------
\begin{figure*}
   \centering
   \includegraphics[width=1\textwidth]{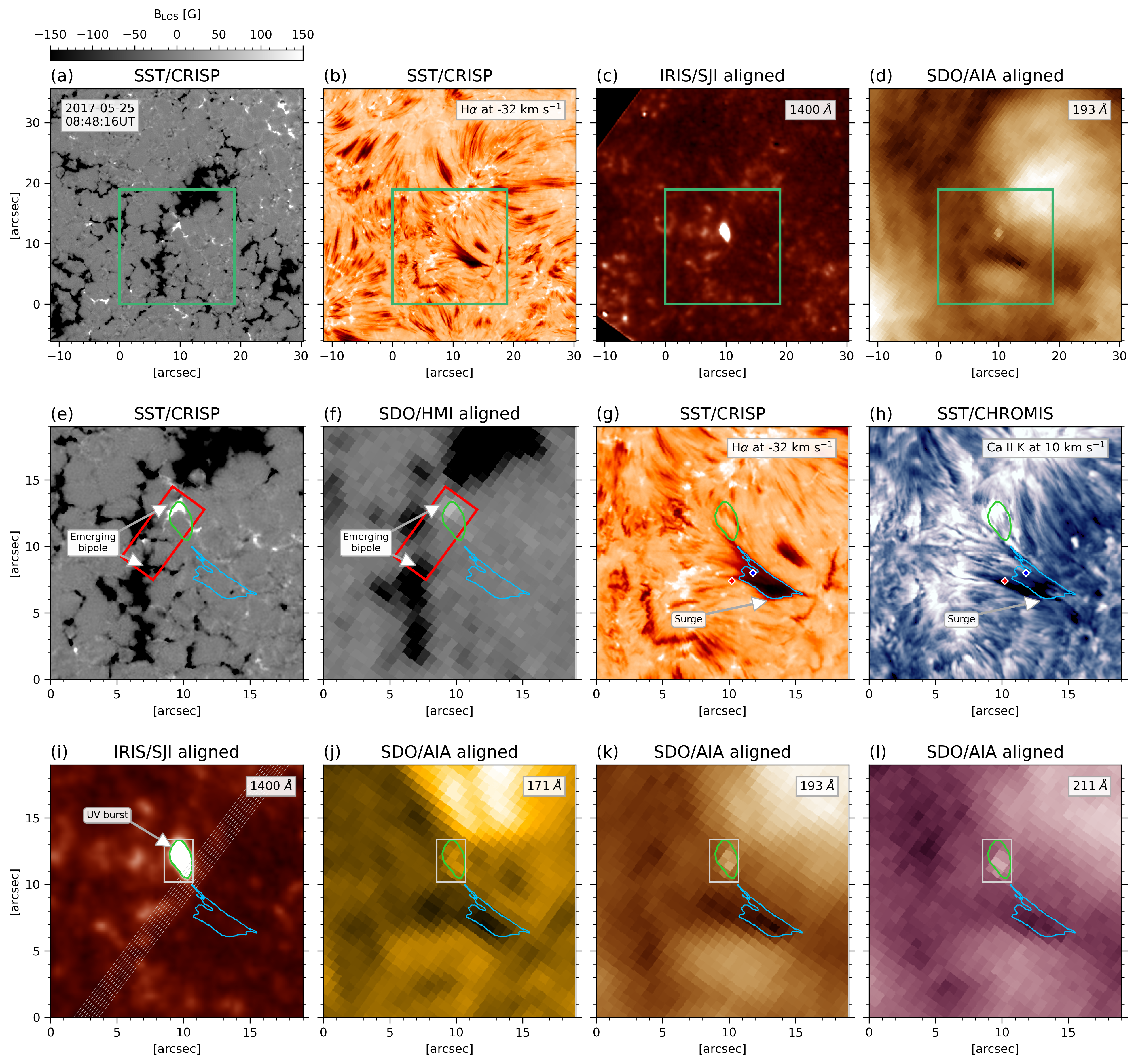}
   \caption{Coordinated observations on 2017 May 25 at $\sim\,$08:48 UT.
   Panel (a): LOS magnetogram for SST/CRISP.
   Panel (b): \Halpha\ at $-32$~km~s$^{-1}$ from SST/CRISP.
   Panel (c): \IRIS/SJI 1400~\AA\ aligned to SST.
   Panel (d): SDO/AIA 193~\AA\ aligned to SST.
   The $19\arcsec\times19\arcsec$ field-of-view highlighted in green encompasses the region of interest
   shown in the rest of the panels. 
   Panels (e) and (f): LOS magnetograms for SST/CRISP and \SDO/HMI aligned to SST, respectively.
   The red rectangle delimits the area where magnetic flux is analyzed in Sect.~\ref{s:flux_emergence}.   
   Panel (g): \Halpha\ at $-32$~km~s$^{-1}$ from SST/CRISP.
   The colored diamonds indicate the location of the spectral profiles shown in Sect.~\ref{s:surge}.
   Panel (h): \CaIIK\ at $10$~km~s$^{-1}$ taken with SST/CHROMIS.
   Panel (i): \IRIS/SJI 1400~\AA\ with 8 lines indicating the approximate locations of the spectrograph slit and the area covered by the raster scan.
   Light curves within the light-gray rectangle are studied in Sect.~\ref{s:uv_burst}.  
   Panels (j), (k), and (l): SDO/AIA 171~\AA, 193~\AA, and 211 \AA, respectively. 
   In all the zoomed-in panels, the colored contours delimit the surge in the \Halpha\ blue wing (blue) 
   and the UV burst on the IRIS/SJI 1400 (green).
   An animation of this figure is available \href{online}{online} with the evolution of the system from 08:15 to 09:02 UT.
   }
   \label{f:figure_01}
\end{figure*}
%---------------------------------------------------------------------------------------------------------------------------

We have used coordinated observations from SST, IRIS, and SDO obtained on 2017 May 25 from 08:07:54 to 09:02:42 UT 
of an enhanced network region near disk center at heliocentric coordinates $(x,y) = (36\arcsec,-91\arcsec)$ 
and observing angle $\mu=0.99$.
This dataset was acquired as part of a multi-season effort to acquire coordinated observations of different targets with SST and IRIS
\citep[see][]{Rouppe-van-der-Voort_etal:2020}. % SST+IRIS database
The observations are unique with a combination of precise co-pointing and excellent seeing conditions over extended period. 
All instruments delivered consistent high-quality data to cover the full evolution of the sequence of events. 
\begin{itemize}
    %------------------------------------------------------------------------------
    \item The SST data set contains the following spectral scans obtained with the 
    CRisp Imaging Spectropolarimeter \citep[CRISP;][]{Scharmer_etal:2008}: 
    the \FeI\ 6301 and 6302~\AA\ line pair in spectropolarimetric mode and \Halpha\ 6563~\AA\ in spectroscopic mode.
    The \FeI\ 6301 and 6302~\AA\ lines were sampled 
    in 16 positions between $-1178$~m\AA\ and $+90$~m\AA\ from the 6302~\AA\ line 
    core. 
    For each line position and polarization state, 6 exposures were acquired that were used for image restoration (i.e., a total of 384 exposures per spectral line scan). 
    The noise level in the restored Stokes V/I$_\textrm{cont}$ maps was estimated to be 2$\times$10$^{-3}$.   
    The \Halpha\ line was sampled in 32 positions 
    ranging from $-1850$~m\AA\ to $+1850$~m\AA\ from the line core. 
    The CRISP scans have a temporal cadence of $19.5$~s, a spatial sampling of 
    0\farcs057 per pixel, and a field of view (FOV) of 
    $\sim 58\arcsec \times 58\arcsec$.

    In addition, the SST data set contains spectral scans from the CHROmospheric 
    Imaging Spectrometer (CHROMIS) for the \CaIIK\ 3934~\AA\ line, which is 
    sampled in 41 positions between $-1300$~m\AA\ and $+1300$~m\AA\ from the 
    line core.
    In addition, one continuum wavelength at 4000~\AA\ was recorded. % 3999.8980 A
    The CHROMIS scans have a temporal cadence of $13.8$~s, a spatial sampling of 
    0\farcs038 per pixel, and a FOV of $66\arcsec \times 42\arcsec$.

    The CRISP and CHROMIS data were processed using the SSTRED data 
    reduction pipeline \citep{de-la-Cruz-Rodriguez_etal:2015a,
    Lofdahl_etal:2021}, which includes Multi-Object Multi-Frame Blind 
    Deconvolution (MOMFBD) \citep{Van-Noort_etal:2005} image restoration and the method for intra-scan consistency of \cite{Henriques:2012}.
    High image quality was achieved with the combination of the SST adaptive optics system \citep{Scharmer_etal:2023}, % Scharmer+ AO system
    MOMFBD image restoration, and excellent seeing conditions. 
    The CRISP and CHROMIS time series were coaligned such that the CRISP data match the CHROMIS pixel scale and the CHROMIS data match the CRISP temporal cadence. 
    More details about the observing program and data processing are provided by \citet{Bose_etal:2021} who analyzed a 97~min time series of the same target that started shortly after the dataset we present here.

    Line-of-sight (LOS) magnetograms were obtained by performing Milne-Eddington 
    inversions of the \FeI\ 6301 and 6302 \AA\ Stokes profiles following the scheme 
    developed by \citet{de-la-Cruz-Rodriguez:2019}\footnote{\url{https://github.com/jaimedelacruz/pyMilne}}. 
    Measuring the standard deviation in a very quiet region,
    we have estimated the noise level in the $B_\mathrm{LOS}$ maps to be 7~G.
    
    %------------------------------------------------------------------------------
    \item The IRIS data set was acquired with the observation program OBSID 3633105426.
    The spectral information was acquired through medium dense raster scans, with the spectrograph slit
    covering $62\arcsec$ in the $y$ direction and scanning 2\arcsec\ in the $x$ 
    direction in 8 steps of 0\farcs35, with an exposure time of 2~s and a raster 
    cadence of 25~s. 
    The slit-jaw images (SJI) covered a FOV of $60\arcsec \times 68\arcsec$ 
    with a cadence of $6$~s, obtaining a total of 4000 images in the 1400~\AA\ 
    and 2796~\AA\ passbands.
    The IRIS data is co-aligned to SST by cross-correlation between \CaIIK\ and 
    SJI 2796~\AA, similarly to \cite{Bose_etal:2019}.
    We estimate that the alignment precision is better than the size of the IRIS pixel \citep[0\farcs17, see][]{Rouppe-van-der-Voort_etal:2020}. 
    %------------------------------------------------------------------------------
    \item We have also employed data from HMI and the \textit{Atmospheric Imaging 
    Assembly} \citep[AIA;][]{Lemen_etal:2012} on board SDO.
    The precision error for 45-second $B_\mathrm{LOS}$ maps is around 10~G \citep[][also see the HMI data documentation\footnote{\url{http://hmi.stanford.edu/Description/hmi-overview/hmi-overview.html}}]{Liu_etal:2012}.
    Regarding the alignment, we used the routines developed by Prof. Rob Rutten and detailed documented in his manual\footnote{\url{https://robrutten.nl/rridl/00-README/sdo-manual.html}}.
    A first cross-alignment was performed in all the 
    AIA channels to the HMI continuum.
    Then, HMI continuum images are co-aligned to \CaIIK\ wideband channels.
    The resulting EUV cross-alignments usually reach up to 0\farcs1  precision.
\end{itemize}

%---------------------------------------------------------------------------------------------------------------------------
\begin{figure*}[!ht]
   \centering
   \includegraphics[width=1\textwidth]{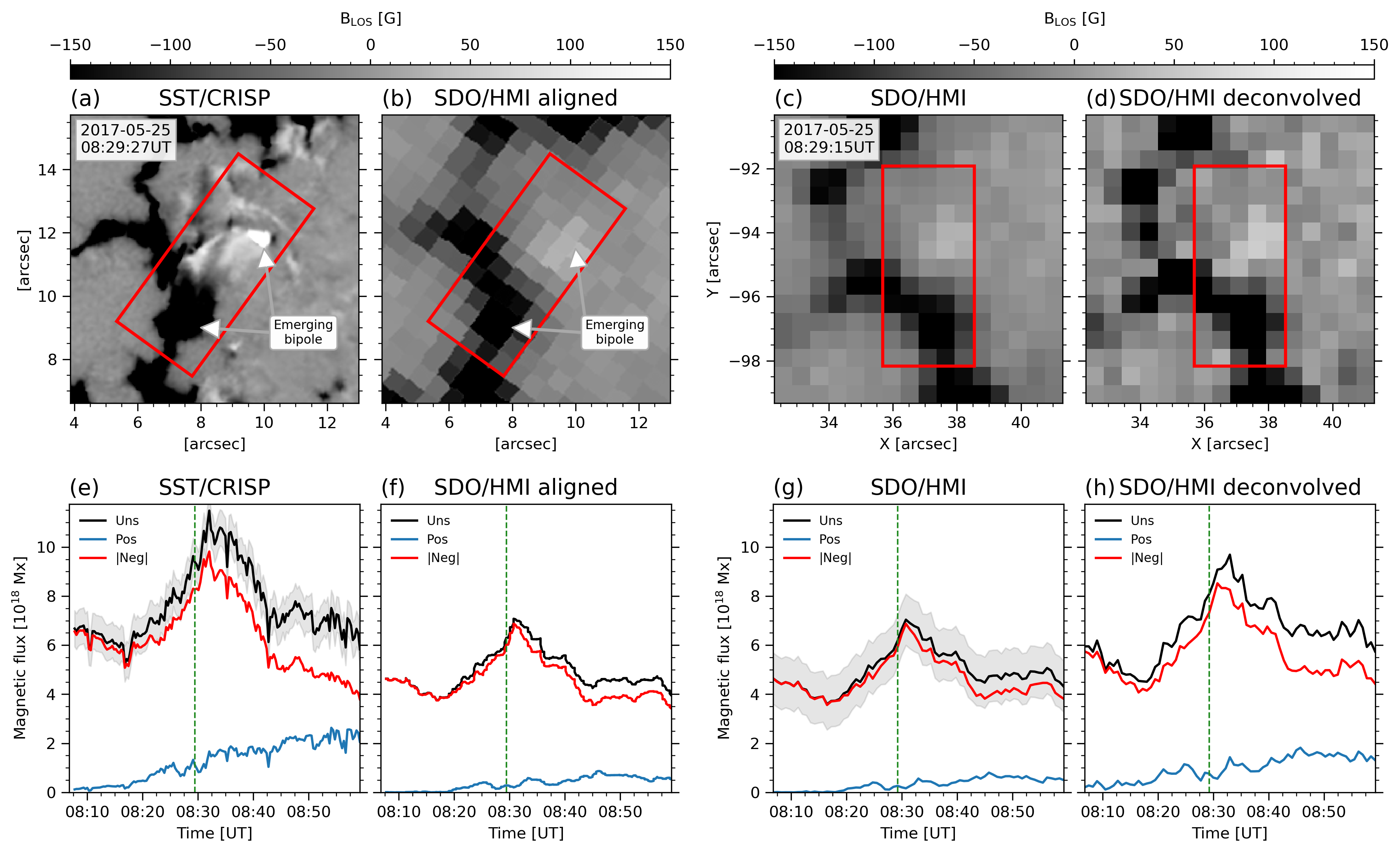}
   \caption{The magnetic flux emergence episode. 
   Panels (a) and (b): LOS magnetograms, respectively, from SST/CRISP and \SDO/HMI aligned to SST.
   Panels (c) and (d): LOS magnetograms from the original \SDO/HMI data without and with deconvolution, respectively.
   Panels (e), (f), (g), and (h): Magnetic flux evolution within the red rectangle for the corresponding magnetograms, 
   showing the unsigned flux (black), the positive flux (blue), and the absolute value of the negative
   flux (red).
   In panels (e) and (g), the gray band represents the standard deviation of the unsigned flux measured in the original SST/CRISP and SDO/HMI data, respectively.
   An animation of this figure is available \href{online}{online} showing the time evolution 
   from 08:07 to 08:59 UT.
   }
   \label{f:figure_02}
\end{figure*}

%---------------------------------------------------------------------------------------------------------------------------
Figure \ref{f:figure_01} and associated animation illustrate the observations on 2017 May 25 analyzed in this paper.
The first row displays the extended context of the observations.
The region of interest is highlighted with a green square and is shown in detail in the bottom two rows.
Panels (e) and (f) contain LOS magnetograms indicating the location of the emerged bipole that leads to
the chain of energetic solar atmospheric events.
In panels (g) and (h), the surge is identifiable as an elongated and dark structure in \Halpha\ and \CaIIK, respectively.
Panel (i) shows \IRIS/SJI 1400 $\AA$ with a UV burst: the compact area with enhanced emission in the middle of the panel around (10\arcsec, 12\arcsec).
Panels (j), (k), and (l) show the counterparts of the UV burst and surge in coronal SDO/AIA channels: the former, as a weak brightening co-located with the UV burst; the latter, as a dark absorption feature in AIA at the same position of the surge.

%%%%%%%%%%%%%%%%%%%%%%%%%%%%%%%%%%%%%%%%%%%%%%%%%%%%%%%%%%%%%%%%%%%%%%%%%%%%%%%%%%%%%%%%%%%%%%%%%%%%%%%%%%%%%%%%%%%%%%%%%%%%
% SECTION 3: RESULTS
%%%%%%%%%%%%%%%%%%%%%%%%%%%%%%%%%%%%%%%%%%%%%%%%%%%%%%%%%%%%%%%%%%%%%%%%%%%%%%%%%%%%%%%%%%%%%%%%%%%%%%%%%%%%%%%%%%%%%%%%%%%%
\section{Results}\label{s:results}
%---------------------------------------------------------------------------------------------------------------------------

%%%%%%%%%%%%%%%%%%%%%%%%%%%%%%%%%%%%%%%%%%%%%%%%%%%%%%%%%%%%%%%%%%%%%%%%%%%%%%%%%%%%%%%%%%%%%%%%%%%%%%%%%%%%%%%%%%%%%%%%%%%%
\subsection{The magnetic flux emergence episode}\label{s:flux_emergence}
%---------------------------------------------------------------------------------------------------------------------------

%---------------------------------------------------------------------------------------------------------------------------
We begin the analysis of the magnetic flux emergence episode by comparing panels (a) and (b) of Fig.~\ref{f:figure_02}, 
which show a close-up of the SST/CRISP and the aligned SDO/HMI magnetograms, respectively.
In the figure, a clear bipole is discernible in the high-resolution (0\farcs057 pixel$^{-1}$) SST/CRISP magnetograms. 
The bipole appears around 08:17 UT (see associated animation) and rapidly shows diverging opposite polarities reaching
a maximum separation around 3\arcsec. 
The bipole is discernible during 20 minutes, approximately, until its negative polarity merges with the preexisting 
surrounding negative field and the positive one starts to collide against the negative environment.
In contrast, the lower resolution SDO/HMI magnetograms (0\farcs5 pixel$^{-1}$) can only detect a slight increase in the 
positive magnetic field across a few pixels.
This makes it challenging to relate it to a clear emerging bipole. 
This result is not altered by the interpolations from the rotation and alignment of the SDO/HMI data 
to SST/CRISP, as demonstrated in panel (c) of Fig.~\ref{f:figure_02}, with the original magnetogram from HMI.
We have further inspected the capabilities of SDO/HMI applying deconvolution following the method of \cite{Norton_etal:2018}, which enhances the SDO/HMI signal by correcting it from scattered 
light. 
The result is shown in panel (d) labeled as SDO/HMI deconvolved.
In this case, the positive magnetic field exhibits higher contrast, but a discernible emerging bipole 
remains elusive.

%---------------------------------------------------------------------------------------------------------------------------
\begin{figure*}
    \centering
    \includegraphics[width=\textwidth]{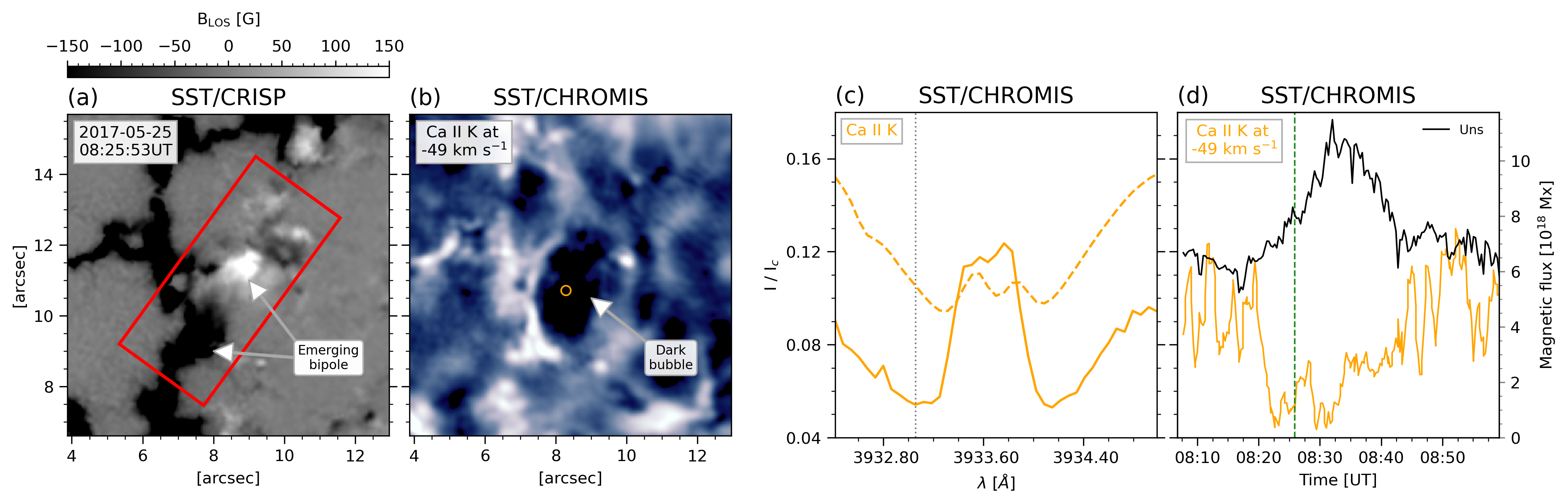}
    \caption{The dark bubble.
    Panel (a): LOS magnetogram from SST/CRISP.
    Panel (b): \CaIIK\ map at $-49$~km~s$^{-1}$ from SST/CHROMIS.
    Panel (c): \CaIIK\ profiles normalized to the continuum for 1) the location marked in Panel (b) (orange solid-line), 
    and 2) the average over the whole FOV of the observation (orange dashed-line).
    Panel (d): Time evolution of 1) the normalized \CaIIK\ intensity at $-49$~km~s$^{-1}$ for the position indicated 
    in panel (b) (orange), and 2) the unsigned magnetic flux from SST/CRISP within the red rectangle (black).
    An animation of this figure is available \href{online}{online} showing the time evolution 
    from 08:07 to 08:59 UT.}
    \label{f:figure_03}
\end{figure*}
%---------------------------------------------------------------------------------------------------------------------------

%---------------------------------------------------------------------------------------------------------------------------
We have delved into the characterization of this event by measuring the magnetic flux within the red 
rectangle depicted in the respective magnetograms. 
The measurements for the unsigned flux (black), positive flux (blue), and absolute value of the
negative flux (red) are shown in the bottom row of Fig.~\ref{f:figure_02}.
In panel (e), the unsigned magnetic flux measured with SST/CRISP is plotted
with its standard deviation (gray band).
The values obtained ($[0.6-1.1]\times 10^{19}$ Mx) are within the ephemeral flux region category \citep{Zwaan:1987} and fit with the flux values reported in other similar emergence episodes \citep[see, e.g.,][]{Guglielmino_etal:2010,JoshiReetika_etal:2017}.
Concerning the positive flux, this shows a linear increase from 08:17 UT onwards as the bipole emerges.
On the contrary, the homologous measurement from the aligned SDO/HMI (panel (f)) does not capture that trend and
underestimates the positive flux up to a factor $\approx4$.
For the negative flux, the underestimation can be up to $1.3$ at the peak of the event at 08:31 UT.
This difference is independent of the rotation and alignment of the SDO/HMI data.
We demonstrate that by measuring the flux in the original data (panel (g)), obtaining the same result.
If we consider the $10$~G precision error specified by the HMI team
for the unsigned flux in the original SDO/HMI data (gray band), we cannot still reproduce the flux measured using SST/CRISP.
The discrepancy in the magnetic flux is noticeably reduced when measuring it in the SDO/HMI deconvolved data 
(panel (h)), although it still underestimates the flux of the event.

%---------------------------------------------------------------------------------------------------------------------------
In addition to the described bipolar emergence episode, 
other secondary smaller bipoles are emerging around the aforementioned main event.
For instance, SST/CRISP magnetograms show a tiny bipole emerging around 08:49 UT centered at (7\farcs0, 10\farcs5) 
which roughly lasts for 5 minutes (see associated animation).
The resolution of SDO/HMI, whereas, does not show any hint about the existence of this event.
Its resolution is not enough to resolve this tiny bipole whose size is around $1$\arcsec.
The appearance of tiny bipoles around a main dominant one in ephemeral flux regions can also be found in numerical
experiments \citep{Nobrega-Siverio_etal:2016}.

%%%%%%%%%%%%%%%%%%%%%%%%%%%%%%%%%%%%%%%%%%%%%%%%%%%%%%%%%%%%%%%%%%%%%%%%%%%%%%%%%%%%%%%%%%%%%%%%%%%%%%%%%%%%%%%%%%%%%%%%%%%%
\subsection{The dark bubble}\label{s:dark_bubble}
%---------------------------------------------------------------------------------------------------------------------------
The effects of magnetic flux emergence become immediately apparent in the lower atmosphere as illustrated 
in Fig. \ref{f:figure_03}.
Panel (a) shows the SST/CRISP magnetogram, providing context, while panel (b), shows a \CaIIK\ map in the blue wing 
at $-49$ km s$^{-1}$.
The latter displays a roundish dark structure, or bubble, situated between the bipole polarities.
In the animation, the dark bubble becomes perceptible around 08:21 UT, four minutes after the emergence 
appears in the photosphere.
Subsequently, the bubble expands in alignment with the diverging motion of the bipole polarities, reaching a maximum 
size of 3\arcsec and lasting for 18 minutes.
The spectral profile at, approximately, the center of the bubble is shown as a solid orange curve in panel (c), 
alongside the average \CaIIK\ profile of the observation (orange dashed curve).
Within the dark bubble, the \CaIIK\ wings exhibit noticeable faintness, about 1.5 times weaker than the average profile,
suggesting a temperature decrease in the lower atmosphere, specifically, a few hundred kilometers above the 
surface. 

%---------------------------------------------------------------------------------------------------------------------------
The relationship between the emergence episode and the appearance of this dark bubble is further evident in panel (d), 
which demonstrates how the intensity of \CaIIK\ at $-49$ km s$^{-1}$ at the center of the bubble diminishes while the magnetic flux increases, that is, they are anti-correlated.
Thus, the appearance of dark bubbles may serve as additional indirect indicators of magnetic flux emergence when high-resolution magnetograms are not available. 
In fact, the secondary tiny bipole described in the previous section also exhibits an associated dark bubble around 
at 08:49 UT centered at  (7\farcs0, 10\farcs5)  (see associated animation).

%---------------------------------------------------------------------------------------------------------------------------%
We have checked the IRIS/SJI 2796 \AA\ images to see if we could discern
a similar dark pattern associated with the bubble like the one reported by \cite{Ortiz_etal:2016}; nonetheless, we have not been able to distinguish any clear high-contrast dark feature.
We have also examined the IRIS \MgII\ h\&k spectra; however, the overlap of the IRIS slit with the bubble is not perfect, so the results are inconclusive.

%---------------------------------------------------------------------------------------------------------------------------
\begin{figure*}[!ht]
    \centering
    \includegraphics[width=\textwidth]{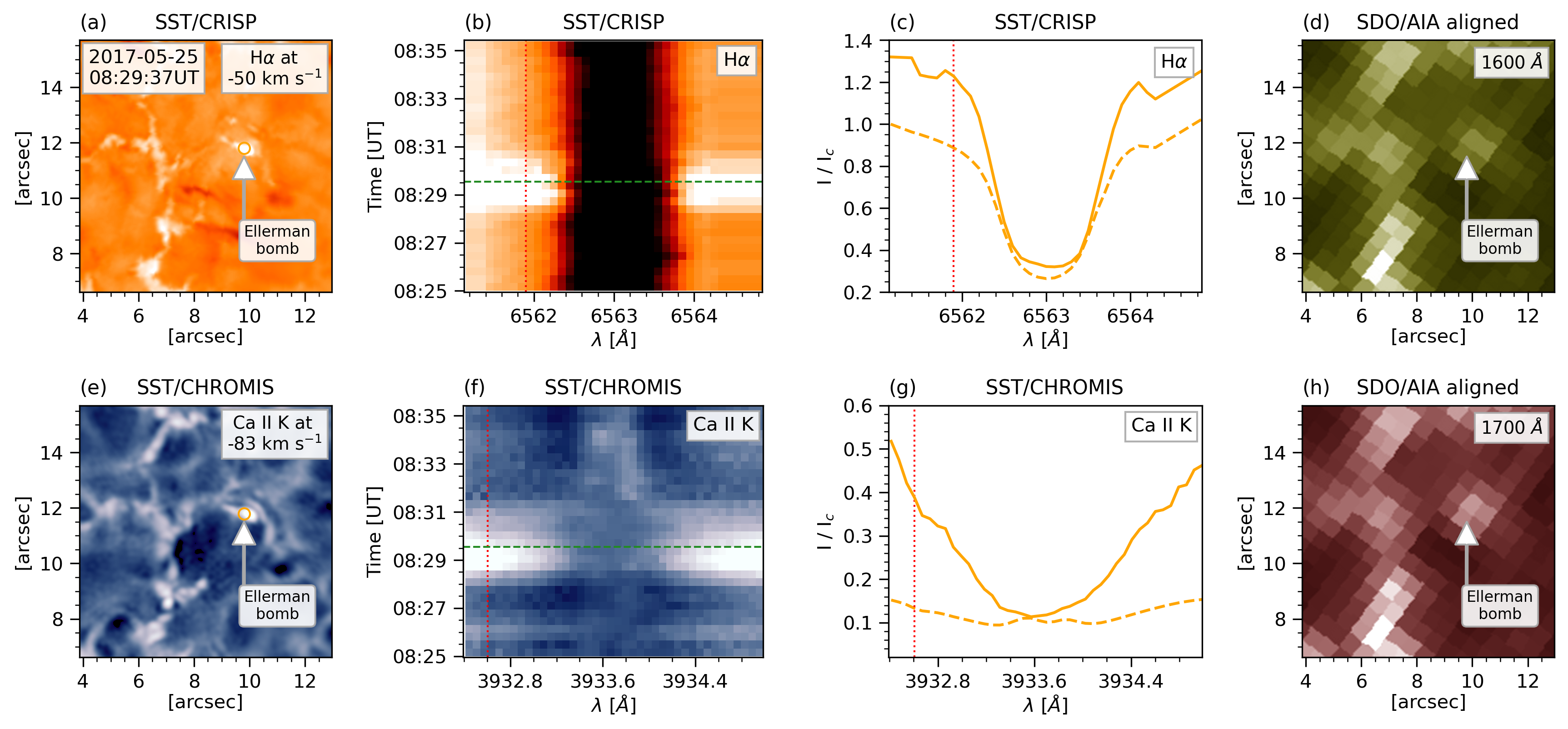}
    \caption{The Ellerman Bomb.
    Panels (a) and (e): Maps of the \Halpha\ and \CaIIK\ lines, respectively, in the far-blue-wing.
    The location of the EB centered at (9\farcs8, 11\farcs8) is indicated with a circle.
    Panels (b) and (f): $\lambda-t$ plot at the center of the EB for \Halpha\ and \CaIIK, respectively.
    The vertical red line indicates the spectral position shown in panels (a) and (e), while the 
    horizontal green line indicates the time.
    Panels (c) and (g): Comparison between the spectral profile of the EB (solid) and the average profile 
    over the whole FOV of the observation (dashed) for \Halpha\ and \CaIIK, respectively.
    Panels (d) and (h): SDO/AIA 1600 and 1700 \AA\ channels aligned to SST.
    An animation of this figure is available \href{online}{online} showing the time evolution 
    from 08:25 to 08:35 UT.
    }
    \label{f:figure_04}
\end{figure*}
%---------------------------------------------------------------------------------------------------------------------------

%%%%%%%%%%%%%%%%%%%%%%%%%%%%%%%%%%%%%%%%%%%%%%%%%%%%%%%%%%%%%%%%%%%%%%%%%%%%%%%%%%%%%%%%%%%%%%%%%%%%%%%%%%%%%%%%%%%%%%%%%%%%
\subsection{The Ellerman bomb (EB)}\label{s:ellerman_bomb}
%---------------------------------------------------------------------------------------------------------------------------
Once the emerging magnetized plasma reaches the lower atmosphere, it interacts with the pre-existing magnetic field, 
giving rise to an EB.
To illustrate it, panels (a) and (e) of Fig.~\ref{f:figure_04} contain far-blue-wing maps of the \Halpha\ and \CaIIK\ lines,
respectively. 
The EB manifests as an enhanced, blob-like brightening in both maps, as pointed out by the arrows.
In the associated animation, we can distinguish its appearance around 08:28 UT, 
which is seven minutes after the initial indications of the dark bubble.
Following the motion of the EB, we can estimate its lifetime to be around 7 min.
During this time, the EB exhibits rapid changes in its apparent shape an size, covering an area up to, approximately, 0.6~arcsec$^2$, which is within the typical values reported in the literature \cite[see, e.g.,][]{Vissers_etal:2013}.
The EB's location, at (9\farcs8, 11\farcs8), coincides with the region where the positive polarity of the emerging
bipole encounters a pre-existing negative concentration: a favorable configuration for magnetic reconnection.
Panels (b) and (f) display the $\lambda-t$ plot at that particular location for  \Halpha\ and \CaIIK, 
respectively. 
Furthermore, panels (c) and (g) show the comparison between the EB's profile (solid) and the average profile over the whole FOV of 
the observation (dashed) for both spectral lines.
In these panels, the enhancement in the wings of both \Halpha\ and \CaIIK\ lines 
is clearly visible, which is a defining feature of EBs
\citep[see, e.g.,][among others cited in the Introduction]{Rouppe-van-der-Voort_etal:2017,Leenaarts_etal:2018,Vissers_etal:2019a},
corroborating the magnetic reconnection scenario.
Additional indications of the EB can be observed in the SDO/AIA 1600 and 1700 \AA\ channels (panels (d) and (h)) 
through a co-spatial slight intensity increase (see associated animation). 
This agrees with previous observational results \citep[see][and references therein]{Vissers_etal:2019a} and further supports the validity of the automated procedure for identifying EBs using SDO data by \cite{Vissers_etal:2019b}.
%
%However, without coordination with other instruments providing high-resolution magnetograms, as in our case, we might overlook the magnetic origin of such small-scale EBs.

%---------------------------------------------------------------------------------------------------------------------------
\begin{figure}
    \centering
    \includegraphics[width=0.5\textwidth]{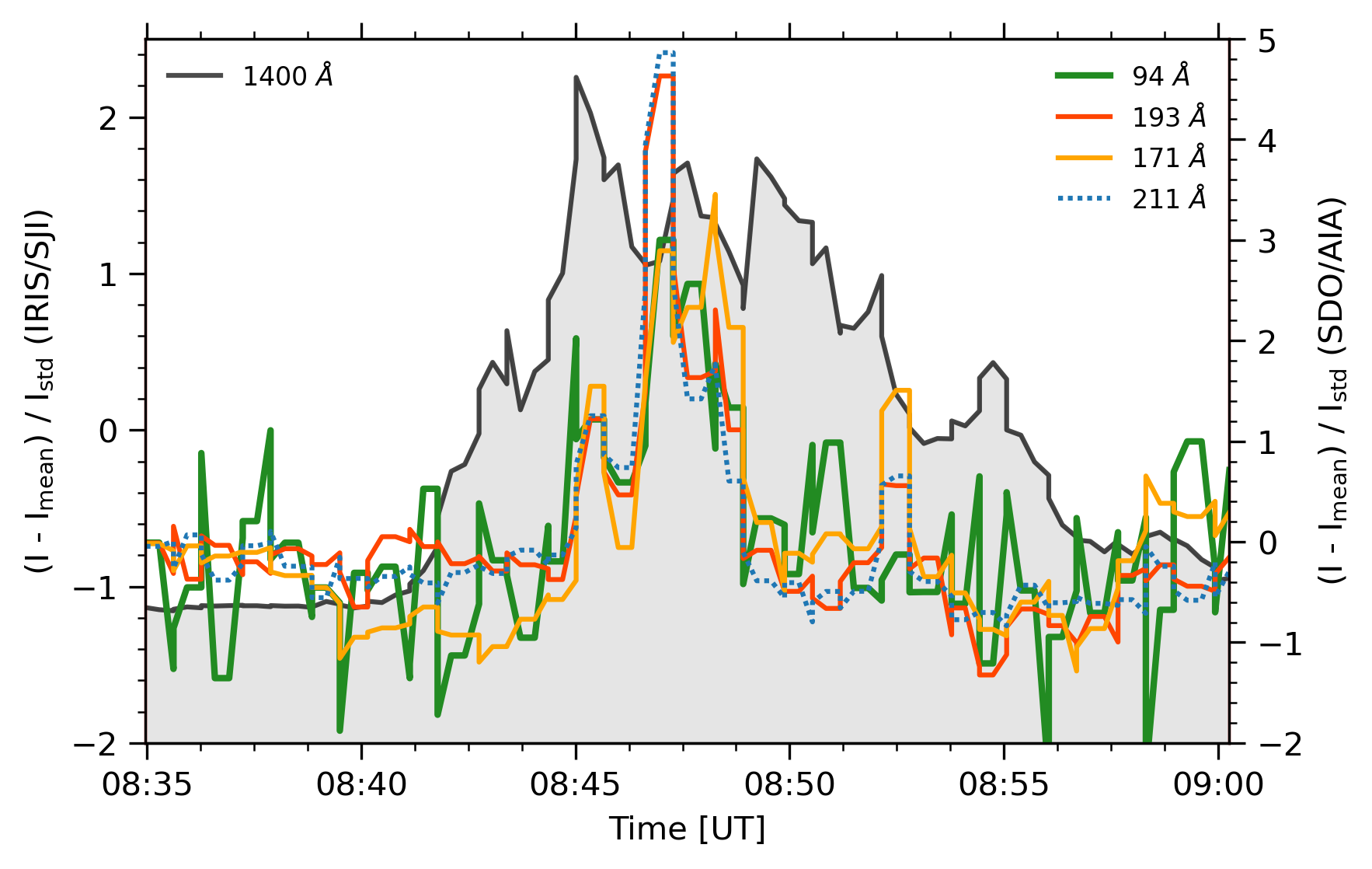}
    \caption{Standardized intensity calculated within the light-gray rectangle shown in Fig.~\ref{f:figure_01}.    
    The curve from IRIS (left axis) shows the UV burst in SJI 1400 \AA\ (black), 
    whereas the curves from SDO (right axis) depict its EUV counterpart in the AIA filters 
    94 \AA\ (green), 193 \AA\ (red), 171 \AA\ (yellow), and 211 \AA\ (blue).
    }
    \label{f:figure_05}
\end{figure}
%---------------------------------------------------------------------------------------------------------------------------

%---------------------------------------------------------------------------------------------------------------------------
\begin{figure*}
    \centering
    \includegraphics[width=1\textwidth]{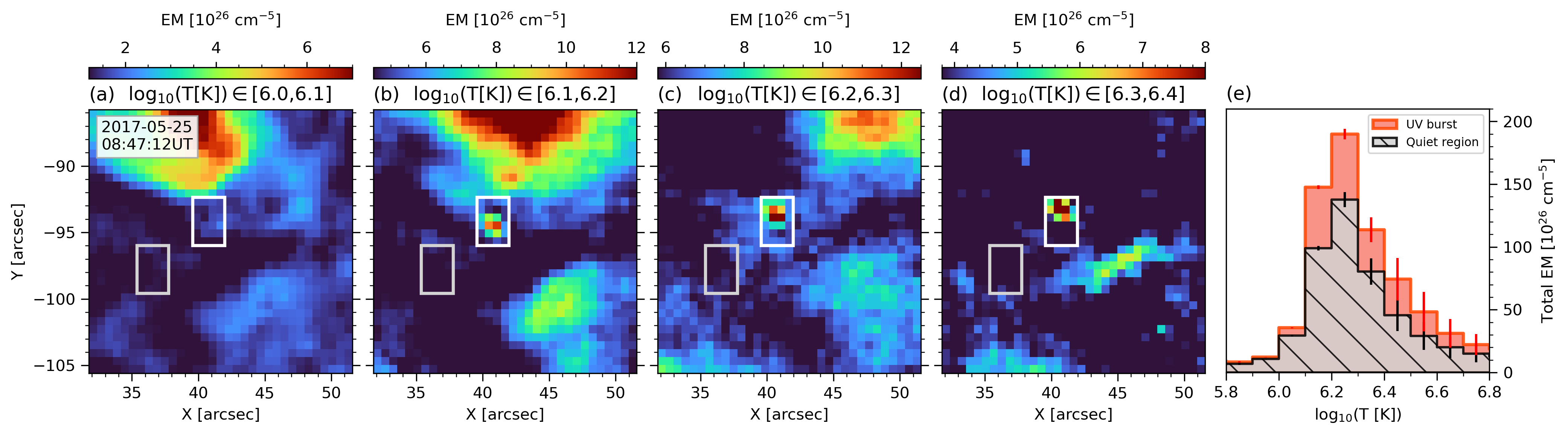}
    \caption{Emission Measure (EM) results applying the method by \cite{Plowman_Caspi:2020} on the original SDO/AIA data 
    at 08:47:12~UT.
    Panels (a)--(d) show the EM maps integrated within the four distinct temperature bins specified in the respective titles.
    The white rectangle centered on $(X,Y) = (41\arcsec, -94\arcsec)$ outlines the EUV counterpart of the UV burst, while the other one in gray delimits a quiet region for comparisons.
    Panel (e) contains the total EM within these two rectangles against temperature, where the associated errors are indicated as vertical lines.
    An animation of this figure is available \href{online}{online} showing the time evolution 
    from 08:35 to 09:00 UT. 
    }
    \label{f:figure_06}
\end{figure*}
%-----------------------------------------------------------------------------------------------------------------------

%%%%%%%%%%%%%%%%%%%%%%%%%%%%%%%%%%%%%%%%%%%%%%%%%%%%%%%%%%%%%%%%%%%%%%%%%%%%%%%%%%%%%%%%%%%%%%%%%%%%%%%%%%%%%%%%%%%%%%%%%%%%
\subsection{The UV burst}\label{s:uv_burst}
%---------------------------------------------------------------------------------------------------------------------------
Additional evidence of magnetic reconnection is found following the EB by the appearance of a UV burst at 08:41~UT 
in the IRIS SJI/1400 \AA\ channel (see Fig.~\ref{f:figure_01} and accompanying animation).
During its lifetime, the UV burst extends up to a width of 2\arcsec\ and of a length of 3\arcsec, approximately.
We have calculated the total intensity $I$ within the light-gray rectangle depicted in Fig.~\ref{f:figure_01}, 
standardizing it, that is, subtracting the average $I_{\mathrm{mean}}$ and dividing it by the standard deviation 
$I_{\mathrm{std}}$.
The result is presented in Fig.~\ref{f:figure_05} as a black line.
The UV burst experiences a rapid increase in intensity over four minutes until it reaches its maximum at 08:45~UT.
Subsequently, the intensity follows a gradual decline, punctuated by sporadic and sudden bursts of intensity enhancements, 
until it disappears around 09:00 UT, resulting in a UV burst lifetime of 19 min.
This bursty behaviour is one of the most characteristic identifying features of UV bursts \citep{Young_etal:2018}.
Interestingly, the UV burst also exhibits a weak EUV counterpart during its lifetime.
This is more evident in the intensity curves of various SDO/AIA channels depicted as colored lines in Fig.~\ref{f:figure_05}.
In the plot, the SDO/AIA 94 \AA\ (green), 193 \AA\ (red), and 211 \AA\ (blue) curves increase immediately   
after the maximum of the UV burst, peaking around 08:47:12 UT, whereas the SDO/AIA 171 \AA\ channel (yellow) shows a bursty behaviour with multiple peaks, 
more akin to the behaviour detected in UV, reaching its maximum intensity slightly later.

To determine whether the SDO/AIA response signifies a bright transition region \citep[see, e.g.,][]{Viall_Klimchuk:2015} or if it reflects a genuine MK response, we have used the approach developed by \cite{Plowman_Caspi:2020} to calculate the differential emission measure (DEM) on the original SDO/AIA data. 
This code solves for the undetermined DEM inversion problem by employing the Tikhonov regularization technique (more commonly known as the ``$L^{2}$ norm''). 
The idea behind this technique is to impose an additional mathematical constraint in the merit function \citep[see equation 14 of ][]{Plowman_Caspi:2020}{}{} that helps to converge the inversion and discards any physical meaningless solutions.
In return, the algorithm guarantees positive DEM solutions.
Panels (a)--(d) of Fig.~\ref{f:figure_06} contain 2D maps centered on the UV burst showing the emission measure (EM) integrated in
bins of $\log_{10}(T[\mathrm{K}])=0.1$ and covering a temperature range of $\log_{10}(T[\mathrm{K}])=[6.0,6.4]$.
This analysis indicates that the increase in the SDO/AIA light curves we could see in Fig.~\ref{f:figure_05} around 08:47:12~UT corresponds
to an enhancement of the emission at temperatures greater than $\log_{10}(T[\mathrm{K}])=6.1$,
which supports the idea that the plasma in the burst is heated above MK temperatures. 
To delve into this, we have calculated the total EM of the area around the UV burst and a quiet reference area, including the error on the recovered EM.
The error is calculated through a Monte Carlo approach \citep{Press_etal:1992} with 100 random realizations of the input AIA data within the noise range.
The EM is then computed for all the different realizations, providing a measure of the errors through the standard deviation of the obtained solutions.
This is a common approach, as implemented in the DEM code by  \cite{Hannah_Kontar:2012}, and it has been widely adopted for studying inversions from AIA observations \citep[see, e.g.,][]{Su_etal:2018,Devi_etal:2022}. 
For a comprehensive overview of errors, refer to \citet{Guennou_etal:2012a,Guennou_etal:2012b} and \citet{Athiray_Winebarger:2024}.
The results are shown in panel (e) of Fig.~\ref{f:figure_06}.
Our analysis reveals that for the range of $\log_{10}(T[\mathrm{K}])=[5.8,6.3]$, the recovered EM is very well constrained.
Such behavior is also consistent with the findings of a recently published study by \cite{Athiray_Winebarger:2024} where the authors use several EM solvers and report that AIA data are most reliable in recovering EM in the range $\log_{10}(T[\mathrm{K}])=[5.8,6.3]$ (see their Figure 5).
Thus, we can strengthen our claims about the plasma in the burst getting heated above MK temperatures.

It is also interesting to note that the EM distributions of the UV burst and the quiet area are similar in terms of the shape and the temperature bin in which they reach their maximum.
The relative EM enhancement at the burst location between $\log_{10}(T[\mathrm{K}])=[6.0,6.6]$ can be an indication that the temperature increase occurs in a higher-density plasma. 
We can estimate the density necessary to explain the enhanced EM at a given temperature bin.
To that end, we suppose that the EM from the burst is a combination of the EM from the quiet region plus an extra optically-thin layer, so 
\begin{equation}
    EM^{\mathrm{extra}} = EM^{\mathrm{burst}} - EM^{\mathrm{quiet}} =  \int{n_\mathrm{e}\, n_\mathrm{H}\, \mathrm{d}l},
\end{equation}
where $n_e$ and $n_H$ are the electron and Hydrogen number density, respectively.
Since the plasma is totally ionized, $n_\mathrm{e} \approx n_\mathrm{H}$, and, if the plasma is uniform in this extra layer, we have
\begin{equation}
    EM^{\mathrm{extra}} = \int{n^2_\mathrm{e}\, \mathrm{d}l} =  n^2_\mathrm{e}\, \int{\mathrm{d}l} = n^2_\mathrm{e}\, L,
\end{equation}
where $L$ is the length along the line-of-sight of the extra layer.
Consequently,
\begin{equation}
    n_\mathrm{e} = \sqrt{\frac{EM^{\mathrm{extra}}}{L}}.
\end{equation}
For instance, $EM^{\mathrm{extra}}=5\times10^{27}$~cm$^{-3}$ for the bin $\log_{10}(T[\mathrm{K}])=[6.2,6.3]$.
Assuming a length $L \in [0.1, 1.0]$~Mm, this would imply 
$n_\mathrm{e} \in [ 0.71,  2.24]\times10^{10}$~cm$^{-3}$, which corresponds to typical densities from the upper chromosphere and transition region.
This is consistent with the UV burst happening below standard coronal heights.
A detailed discussion regarding the EUV counterpart and the MK response is provided in Sect.~\ref{s:discussion}.

%%%%%%%%%%%%%%%%%%%%%%%%%%%%%%%%%%%%%%%%%%%%%%%%%%%%%%%%%%%%%%%%%%%%%%%%%%%%%%%%%%%%%%%%%%%%%%%%%%%%%%%%%%%%%%%%%%%%%%%%%%%%
\subsection{The surge}\label{s:surge}
%---------------------------------------------------------------------------------------------------------------------------
Simultaneously with the UV burst, a chromospheric surge is ejected: an additional indication 
of magnetic reconnection.
The surge is visible in the wings of \Halpha\ and \CaIIK\ (see Fig.~\ref{f:figure_01} and
associated animation) and reaches a maximum length and width of 8\arcsec\ and 3\arcsec, respectively. 
The velocity inferred is around a few tens of km s$^{-1}$ and its lifetime is, approximately, 19 min.
The surge also shows a dark counterpart in all the SDO/AIA EUV channels, indicating that
the surge thickness along the LOS has to be at least of a few Mm to attenuate the AIA emission,
as stated by \cite{Ortiz_etal:2016}.

\begin{figure}
    \centering
    \includegraphics[width=0.5\textwidth]{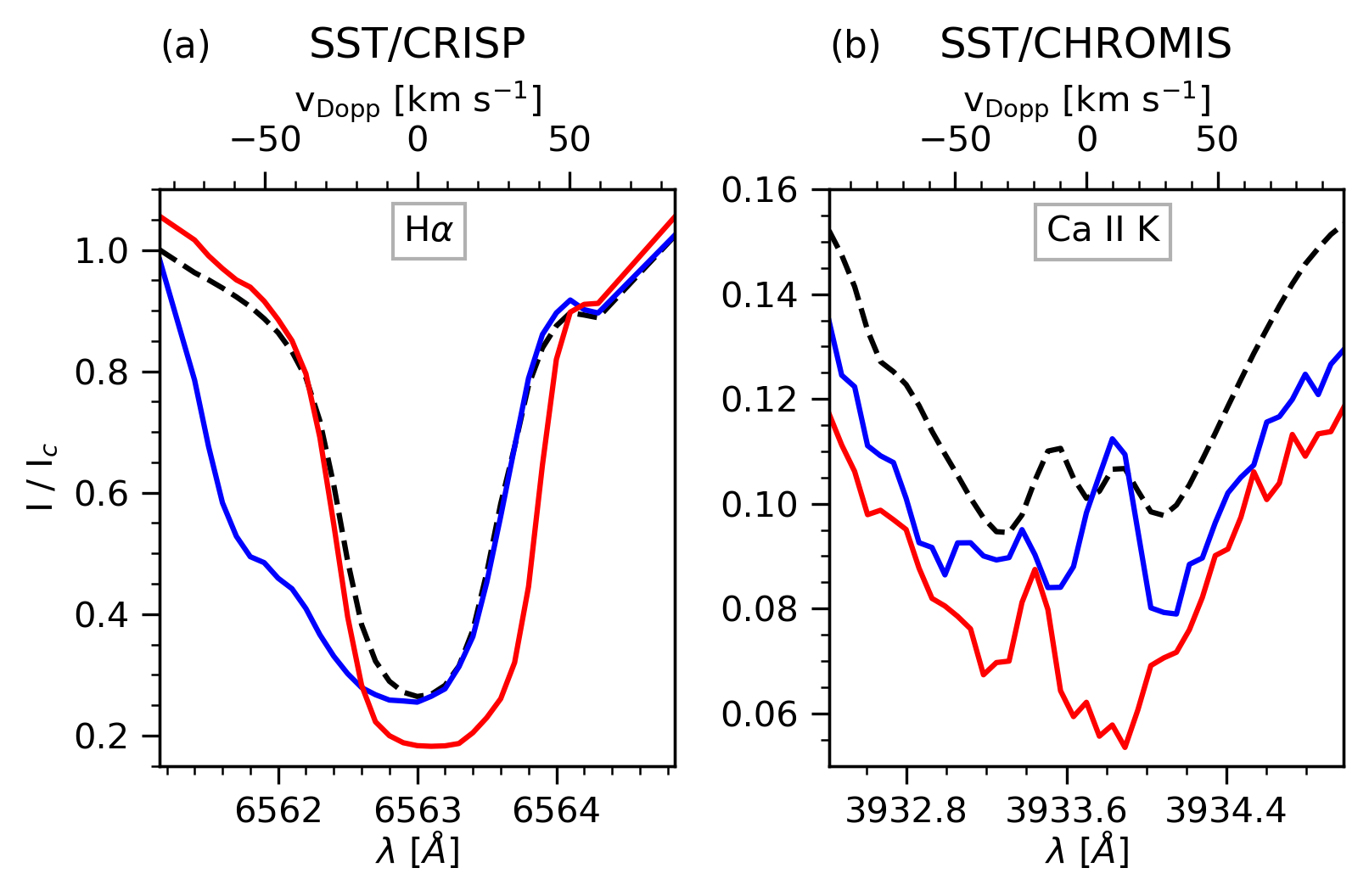}
    \caption{Surge representative spectra for \Halpha\ (panel (a)) and \CaIIK\ (panel (b))
    at $\sim\,$08:48 UT. 
    The blue and red curves are the profiles corresponding to the locations indicated, respectively, 
    by the blue and red diamonds in Fig.~\ref{f:figure_01}.  
    The black dashed line is a reference profile that is the average over the whole FOV of the observation.
    }
    \label{f:figure_07}
\end{figure}

%---------------------------------------------------------------------------------------------------------------------------
We have inspected representative spectral profiles within the surge in \CaIIK\ and relate them with 
the ones obtained in \Halpha.
The chosen locations to study the surge spectra are indicated in Fig.~\ref{f:figure_01} by
colored symbols: a blue diamond for surge plasma ascending and 
a red one for surge plasma descending.
The results are shown in Fig.~\ref{f:figure_07} following the same color code, including,
as a black line, the average profile over the whole FOV for comparison purposes.
Panel (a) shows that the \Halpha\ profiles are standard for a surge, namely, they show blue and red shift of the overall line profile, as well as asymmetry with enhanced absorption in one of the wings, in accordance to the motion of the plasma along the LOS.
%blue and red shifted absorption features  with respect to the average profile, accordingly to the motion of the plasma along the LOS.
%
At the same location in \CaIIK, panel (b),  the ascending surge plasma shows an absent 
K$_\mathrm{2V}$ peak,
while the descending one has an absent K$_\mathrm{2R}$ peak. 
We attribute this absence to the shift of the high opacity \CaIIK\ core (K$_3$), which nearly completely blocks the photons coming from lower regions in the atmosphere that form the K$_\mathrm{2}$ features. 
This matches with the opacity window effect described for other chromospheric ejections such as spicules \citep{Bose_etal:2019}.
%
%It is the same reason why, in radiative transfer, complete redistribution is typically a good approximation for the very core of the line (Scharmer1986 or so) but not for the inner wings. In other words, the photons escape primarily via redistribution to other wavelengths, including the K$_\mathrm{2}$ features and line wings. This redistribution may contribute to the visible higher intensity of the longer wavelength wing of the red profile.
%
%The K$_3$ shifts found in the surge, which are in accordance with the \Halpha\ core Doppler shifts, are an additional evidence that K$_3$ is a good tracer of the LOS velocities in the  chromosphere  \citep[see][]{Bjorgen_etal:2018}. 
%
The inferred K$_3$ shifts, as well as the lower intensity K$_3$ line minimum for the red profile, are in accordance with the corresponding \Halpha\ core Doppler shifts and relative intensities. 
This serves as a direct confirmation that the \ion{Ca}{ii}~K$_3$ feature is a good tracer of the LOS velocities in the chromosphere  \citep[see][]{Bjorgen_etal:2018}. 

%---------------------------------------------------------------------------------------------------------------------------
%---------------------------------------------------------------------------------------------------------------------------

%%%%%%%%%%%%%%%%%%%%%%%%%%%%%%%%%%%%%%%%%%%%%%%%%%%%%%%%%%%%%%%%%%%%%%%%%%%%%%%%%%%%%%%%%%%%%%%%%%%%%%%%%%%%%%%%%%%%%%%%%%%%
% SECTION 4: DISCUSSION  AND CONCLUSIONS
%%%%%%%%%%%%%%%%%%%%%%%%%%%%%%%%%%%%%%%%%%%%%%%%%%%%%%%%%%%%%%%%%%%%%%%%%%%%%%%%%%%%%%%%%%%%%%%%%%%%%%%%%%%%%%%%%%%%%%%%%%%%
\section{Discussion and conclusions}\label{s:discussion}
%---------------------------------------------------------------------------------------------------------------------------
In this paper, we have used coordinated observations from SST, IRIS, and SDO obtained on 2017 May 25 to study  
the whole evolution of an ephemeral magnetic flux emergence episode and subsequent related energetic phenomena at 
the different solar atmospheric layers. 
In the following, we present the discussion and main conclusions for each of these phenomena from Sect.~\ref{s:results}: the emerging bipole, the dark bubble, the EB, the UV burst, and the surge.

%%%%%%%%%%%%%%%%%%%%%%%%%%%%%%%%%%%%%%%%%%%%%%%%%%%%%%%%%%%%%%%%%%%%%%%%%%%%%%%%%%%%%%%%%%%%%%%%%%%%%%%%%%%%%%%%%%%%%%%%%%%%
\subsection{Magnetic flux emergence}
%---------------------------------------------------------------------------------------------------------------------------
The SST/CRISP magnetograms clearly reveal the appearance of a main bipole at the solar surface with a few arcsec of size 
as well as other smaller bipoles emerging in the neighborhood.
In contrast, the SDO/HMI magnetograms barely allow us to distinguish the positive part of the main bipole, substantially underestimating the amount of magnetic flux, even when incorporating the error deviation in the measurements.
Deconvolving the SDO/HMI data \citep{Norton_etal:2018} helps to get a slightly better agreement in the flux values and higher contrast in the magnetograms, but it is still not enough to clearly discern the main emerging bipole, nor the smaller secondary magnetic flux emergence episodes that occur in the surroundings.

%---------------------------------------------------------------------------------------------------------------------------
As a conclusion, our results showcase the importance of employing high-resolution magnetograms to detect 
small-scale magnetic flux emergence episodes and relate them to subsequent eruptive/ejective phenomena. 
Thus, we warn about the use of HMI magnetograms to draw conclusions about the origin of small-scale phenomena in the solar atmosphere.
For instance, the claims about the lack of evidence of flux emergence leading to some surges \citep{Nelson_Doyle:2013} 
or related to the so-called campfires \citep{Panesar_etal:2021,Panesar_etal:2023} could be biased because of insufficient resolution.
We estimate therefore that to properly resolve a bipole like the main one studied in this paper, it is necessary to have 
a resolution, of at least 4 to 5 %, $4-5$ 
times larger than that of HMI.
This could be achievable, for example, using the Polarimetric and Helioseismic Imager \citep[PHI;][]{Solanki_etal:2020} 
onboard SolO, which will reach 200~km resolution at a distance of 0.28 AU; or 
through ground-based telescopes with at least 1~m aperture.

%%%%%%%%%%%%%%%%%%%%%%%%%%%%%%%%%%%%%%%%%%%%%%%%%%%%%%%%%%%%%%%%%%%%%%%%%%%%%%%%%%%%%%%%%%%%%%%%%%%%%%%%%%%%%%%%%%%%%%%%%%%%
\subsection{The dark bubble}
%---------------------------------------------------------------------------------------------------------------------------
To the best of our knowledge, this is the first report in the \CaIIK\ line of the existence of cold domes related
to flux emergence through the appearance of a dark bubble located within the emerging bipole, and it adds to the already 
existing reports in \CaIIH\ \citep{Vargas-Dominguez_etal:2012,Centeno_etal:2017,Kontogiannis_etal:2020},
also referred to as dark lanes \citep{Otsuji_etal:2007,Cheung_etal:2008,Guglielmino_etal:2010}, 
and in the infrared \CaII\ 8542 \AA\ line \citep{Ortiz_etal:2014,de-la-Cruz-Rodriguez_etal:2015b,Ortiz_etal:2016}.

The reason for such intensity dimming in \CaII\ can be explained by adiabatic cooling that takes place when the emerging magnetized plasma rapidly expands as it rises through the solar atmosphere, as shown by \cite{de-la-Cruz-Rodriguez_etal:2015b}.
Even though our results are not conclusive for \MgII\ h\&k due to the location of the IRIS slit with respect to the bubble, it is encouraging to
keep looking for evidence since these lines are formed higher in the solar atmosphere compared to the homologous 
\CaII\ H\&K lines as well as the \CaII\ 8542 \AA\ line \citep{Leenaarts_etal:2013a,Leenaarts_etal:2013b,Bjorgen_etal:2018}, and they could potentially track later stages of the expansion.
Additionally, it would be of interest to explore other diagnostic methods that can provide 
information about cool temperatures in a similar fashion as \cite{Mulay_Fletcher:2021}, using IRIS observations to 
find regions with H$_2$ molecules next to flares, or as \cite{Stauffer_etal:2022},
employing observations in the 4.7 $\mu$m molecular band of CO with the NSO Array Camera 
\citep[NAC;][]{Ayres_etal:2008} to report cool features around a pore.

%%%%%%%%%%%%%%%%%%%%%%%%%%%%%%%%%%%%%%%%%%%%%%%%%%%%%%%%%%%%%%%%%%%%%%%%%%%%%%%%%%%%%%%%%%%%%%%%%%%%%%%%%%%%%%%%%%%%%%%%%%%%
\subsection{The Ellerman bomb (EB)}
%---------------------------------------------------------------------------------------------------------------------------
The reported EB following flux emergence appears to exhibit a very canonical pattern, with enhanced wings in both 
\Halpha\ and \CaIIK, while the core of these lines remains relatively unperturbed.
There is a 7-minute delay between the EB and the appearance of the dark bubble, and an 11-minute delay with respect to the first detection of the bipole. 
These timescales align with findings from 3D MHD numerical simulations, which range from a few minutes
\citep{Danilovic:2017,Hansteen_etal:2017} to several tens of minutes
\citep{Danilovic_etal:2017,Hansteen_etal:2019}.
The delay is attributed to the time the emerged loops need to reach a few hundred kilometers above the surface and create a significant conflict with the preexisting magnetic field in order to lead magnetic reconnection and plasma heating.

Recent observations using the \Hbeta\ line suggest that EBs are more common than previously believed, opening up 
new possibilities for the study of these small-scale reconnection events \citep{JoshiJayant_etal:2020,
JoshiJayant_etal:2022}.
The exact scenario to explain the high occurrence of EBs found in Quiet Sun (QS) regions remains unclear.
Therefore, investigating the relationship between these Quiet Sun Ellerman Bombs (QSEBs) and, for instance, 
flux emergence episodes as small as the secondary episodes reported here, is a task that must be addressed 
in the near future: a challenge that can only be tackled with the availability of high-resolution magnetograms.

%%%%%%%%%%%%%%%%%%%%%%%%%%%%%%%%%%%%%%%%%%%%%%%%%%%%%%%%%%%%%%%%%%%%%%%%%%%%%%%%%%%%%%%%%%%%%%%%%%%%%%%%%%%%%%%%%%%%%%%%%%%%
\subsection{The UV burst}
%---------------------------------------------------------------------------------------------------------------------------
The UV burst following the EB displays canonical features, such as its size, its roundish shape and enhanced intensity with rapid temporal variation. 
Given the debate about the temperature and height formation of UV bursts \citep[e.g,][and references therein]{Judge:2015,Grubecka_etal:2016,Peter_etal:2019,Ni_etal:2022}, the finding of associated weak EUV signatures in the SDO/AIA coronal channels is particularly intriguing.
Through DEM analysis, we show that the EUV response indicates heating above 1 MK temperatures. 
Additionally, the density estimated to explain the enhanced EUV related to the UV burst corresponds to densities between the upper chromosphere and transition region.
These results are consistent
with recent radiative-MHD numerical models of magnetic flux emergence. 
There, UV bursts are formed in elongated current sheets spanning several scale heights through the chromosphere and the plasma is at times heated to above 1 MK, \citep[see, e.g.,][]{Nobrega-Siverio_etal:2017,Rouppe-van-der-Voort_etal:2017, Hansteen_etal:2019}.
From observations, it seems that there is a continuum of EUV counterparts concerning UV bursts: starting with 
\cite{Peter_etal:2014} and \cite{Kleint_Panos:2022}, who did not find any counterparts in AIA images; then moving on to, 
for instance, \cite{Vissers_etal:2015} and \cite{Gupta_Tripathi:2015}, who only detected brief signatures in the 171 and 
193 \AA\ filters; and finally to \cite{Guglielmino_etal:2018}, who observed a very strong counterpart in the SDO/AIA coronal channels
related to the UV burst, even reporting emission in the weak coronal \FeXII\ 1349.4 \AA\ line using IRIS \citep{Guglielmino_etal:2019}.
The diversity of EUV responses may stem from differences in the amount of magnetic energy involved during the flux 
emergence process and subsequent reconnection, as well as differences in the density of the inflow plasma at the 
reconnection site, which determines how efficient the Joule heating per particle is.
Additionally, overlying cool and dense plasma to the reconnection site can also significantly absorb the EUV emission source,
even blocking it completely \cite[see][and their Figure 7]{Hansteen_etal:2019}.

%---------------------------------------------------------------------------------------------------------------------------
Usually, in flux emergence experiments, there is a coronal inverted-Y shaped (or Eiffel-tower) jet with a narrow spine next to the surge once the plasma is heated to temperatures around MK \citep[e.g.,][]{Yokoyama_Shibata:1995,Moreno-Insertis_Galsgaard:2013,Fang_etal:2014,Nobrega-Siverio_etal:2016,Nobrega-Siverio_Moreno-Insertis:2022}.
In our observations, we did not detect such an elongated hot jet next to the surge. 
We cannot exclude that this is due to the resolution limitations of SDO/AIA and that we did not sufficiently spatially resolve the fine structure of this event.
Nonetheless, we have seen that the bursty enhancements in the region of the UV burst detected in the SDO/AIA coronal channels occur on scales as small as 60~s.
These timescales align with those recently reported by \cite{Chitta_etal:2023} concerning tiny inverted-Y-shaped coronal jets with picoflare energy observed using SolO/EUI-HRI.
It is likely that with higher resolution, some bursts could be associated with tiny EUV inverted-Y-shaped jets.

%%%%%%%%%%%%%%%%%%%%%%%%%%%%%%%%%%%%%%%%%%%%%%%%%%%%%%%%%%%%%%%%%%%%%%%%%%%%%%%%%%%%%%%%%%%%%%%%%%%%%%%%%%%%%%%%%%%%%%%%%%%%
\subsection{The surge}
%---------------------------------------------------------------------------------------------------------------------------
The surge ejected as a result of magnetic flux emergence studied here shows typical values with respect to its properties.
For instance, the surge has a lifetime around 19 min, which fits with the values, between several minutes and tens of minutes, that have been derived from both observations \citep[e.g.,][]{Wang_etal:2014,Ortiz_etal:2020,Guglielmino_etal:2018} and simulations \cite[e.g.,][]{Nobrega-Siverio_etal:2016,Li_etal:2023}.
The velocities, around a few tens of km s$^{-1}$ inferred from the shifts in the \Halpha\ and \CaIIK\ lines, are also in agreement with
the moderate velocities that have been carefully analyzed in numerical experiments \citep[see][]{MacTaggart_etal:2015}.

The fact that this surge is very canonical makes it a case study to start exploring their \CaIIK\ properties.
The motivation is that most of the observational surge studies have focused on \Halpha, and the ones studying the violet \CaII\ lines have been traditionally performed with broadband filters centered in \CaIIH, for instance, using Hinode \citep{Nishizuka_etal:2008,Liu_etal:2009,Liu_etal:2011}.
Consequently, the details about the \CaIIK\ spectra have not been explored.
In this paper, we have made a first step to alleviate this, showing the \CaIIK\ spectra in two representative locations
of the surge. 
We find that the blue and red shifts seen in \Halpha\ in the surge have a corresponding counterpart as shifts of the \CaIIK\ core (K$_3$) with same Doppler velocities. 
This can potentially facilitate the surge identification using machine learning techniques in the near future, following approaches like $k$-means \citep[see, e.g.,][]{Nobrega-Siverio_etal:2021}.

\subsection{Concluding remarks}
We have studied a unique dataset of coordinated observations between ground-based and space-borne telescopes covering the full evolution of an ephemeral magnetic flux emergence episode that triggers a chain of subsequent energetic phenomena higher in the atmosphere and ends with an eruption of plasma in the form of a surge. 
The high-resolution SST observations reveal in detail the emergence of a bipole that is practically impossible to discern in the cotemporal HMI magnetograms, emphasizing  the need of high-resolution magnetograms to unveil the magnetic origin of small-scale atmospheric events.
The emerging magnetic flux leads to the formation of a dark bubble, which is an imprint of the passage of the magnetized plasma traversing the low solar atmosphere, and that is firstly reported here in the \CaIIK\ line.
The observations also show the occurrence of magnetic reconnection once the emerging magnetized plasma interacts with the preexisting magnetic field, as evidenced first by the appearance of an EB, and then by a UV burst and surge, both with EUV counterparts. 
This provides an excellent scenario to confront recent numerical MHD simulations.
The EUV emission in the UV burst indicates heating to temperatures above 1~MK in the region of the upper chromosphere and transition region.  
We expect that forthcoming observations from SolO/EUI-HRI will provide valuable insights in this regard.

%%%%%%%%%%%%%%%%%%%%%%%%%%%%%%%%%%%%%%%%%%%%%%%%%%%%%%%%%%%%%%%%%%%%%%%%%%%%%%%%%%%%%%%%%%%%%%%%%%%%%%%%%%%%%%%%%%%%%%%%%%%%
% Acknowledgements
%%%%%%%%%%%%%%%%%%%%%%%%%%%%%%%%%%%%%%%%%%%%%%%%%%%%%%%%%%%%%%%%%%%%%%%%%%%%%%%%%%%%%%%%%%%%%%%%%%%%%%%%%%%%%%%%%%%%%%%%%%%%
\begin{acknowledgements}
This research has been supported by the Spanish Ministry of Science, Innovation, and Universities through projects 
AYA2014-55078-P and PGC2018-095832-B-I00; 
by the European Research Council through the Synergy Grant number 810218 (ERC-2018-SyG); 
by the International Space Science Institute (ISSI) in Bern, through ISSI International Team project \#535
\textit{Unraveling surges: a joint perspective from numerical models, observations, and machine learning}; and
by the Research Council of Norway through its Centres of Excellence scheme, project number 262622, and through project number 325491.
S.B. gratefully acknowledges support from NASA grant 80NSSC20K1272 “Flux emergence and the structure, dynamics, and energetics of the solar atmosphere” and from NASA contract NNG09FA40C (IRIS).
C.F. acknowledges funding from the CNES.
% SST
The Swedish 1-m Solar Telescope is operated on the island of La
Palma by the Institute for Solar Physics of Stockholm University in the Spanish
Observatorio del Roque de Los Muchachos of the Instituto de Astrof\'isica
de Canarias. The Institute for Solar Physics is supported by a grant for research
infrastructures of national importance from the Swedish Research Council (registration
number 2017-00625). 
% IRIS
IRIS is a NASA small explorer mission developed and operated by LMSAL with mission operations executed at NASA Ames
Research center and major contributions to downlink communications funded
by ESA and the Norwegian Space Centre. 
% SDO
SDO observations are courtesy of NASA/SDO and the AIA, EVE, and HMI science teams.
We acknowledge the HMI team for providing the data corrected for scattered light, with special thanks to Dr. Aimee Norton and Dr. Jeneen Sommers for their assistance.
\end{acknowledgements}

%%%%%%%%%%%%%%%%%%%%%%%%%%%%%%%%%%%%%%%%%%%%%%%%%%%%%%%%%%%%%%%%%%%%%%%%%%%%%%%%%%%%%%%%%%%%%%%%%%%%%%%%%%%%%%%%%%%%%%%%%%%%
% BIBLIOGRAPHY
%%%%%%%%%%%%%%%%%%%%%%%%%%%%%%%%%%%%%%%%%%%%%%%%%%%%%%%%%%%%%%%%%%%%%%%%%%%%%%%%%%%%%%%%%%%%%%%%%%%%%%%%%%%%%%%%%%%%%%%%%%%%
\bibliographystyle{aa}
\bibliography{bibliography}

%%%%%%%%%%%%%%%%%%%%%%%%%%%%%%%%%%%%%%%%%%%%%%%%%%%%%%%%%%%%%%%%%%%%%%%%%%%%%%%%%%%%%%%%%%%%%%%%%%%%%%%%%%%%%%%%%%%%%%%%%%%%
% END
%%%%%%%%%%%%%%%%%%%%%%%%%%%%%%%%%%%%%%%%%%%%%%%%%%%%%%%%%%%%%%%%%%%%%%%%%%%%%%%%%%%%%%%%%%%%%%%%%%%%%%%%%%%%%%%%%%%%%%%%%%%%
\end{document}